\documentclass[final,3p]{elsarticle}

\usepackage{epsfig}

\usepackage{amssymb}
\usepackage{hyperref}
\hypersetup{
    colorlinks=true,       % false: boxed links; true: colored links
    linkcolor=red,          % color of internal links
    citecolor=green,        % color of links to bibliography
    filecolor=magenta,      % color of file links
    urlcolor=cyan           % color of external links
}

\def\Journal#1#2#3#4{{#1} {\bf#2} (#4) #3}
\def\EPJA{{Eur. Phys. Jour.} A}

\def\NPA{{ Nucl. Phys.} A}

\def\PLB{{ Phys. Lett.} B}

\def\PRL{ Phys. Rev. Lett.}

\def\PREP{ Phys. Rep.}

\def\PRC{{ Phys. Rev.} C}

\def\INT{{ Int. J. Mod. Phys.} E}

\newcommand{\be}{\begin{equation}}
\newcommand{\ee}{\end{equation}}
\newcommand{\bea}{\begin{eqnarray}}
\newcommand{\eea}{\end{eqnarray}}

\journal{Physics Letters B}

\begin{document}

\begin{frontmatter}

\title{Neutron-proton elliptic flow difference as a probe for 
the high density dependence of the symmetry energy }
%\title{
%Neutron-proton differential squeeze-out as a probe of the isospin dependent part of the equation
%of state of nuclear matter}
\author{M.D. Cozma}
\ead{cozma@niham.nipne.ro}
\address{IFIN-HH, Atomi\c{s}tilor 407, 077125 M\v{a}gurele-Bucharest, Romania\\}

\begin{abstract} 
We employ an isospin dependent version of the QMD transport model to study the influence of the isospin 
dependent part of the nuclear matter equation of state and in-medium nucleon-nucleon
cross-sections on the dynamics of heavy-ion collisions at intermediate energies. We find
that the extraction of useful information on the isospin-dependent part of the equation of state of 
nuclear matter from proton or neutron elliptic flows is obstructed by their sensitivity to model parameters
and in-medium values of nucleon-nucleon cross-sections. Opposite to that, neutron-proton elliptic flow
difference shows little dependence on those variables while its dependence on the isospin asymmetric EoS
is enhanced, making it more suitable for a model independent constraining of the high-density behaviour
of asy-EoS. Comparison with existing experimental FOPI-LAND neutron-hydrogen data can be used to set an upper limit
to the softness of asy-EoS. Successful constraining of the asy-EoS via neutron-proton elliptic flow difference
will require experimental data of higher accuracy than presently available.
\end{abstract}


\end{frontmatter}

\section{Introduction}
\label{introduction}

One of the remaining opened questions in nuclear physics is the equation of state (EoS) of isospin asymmetric
nuclear matter, $\it i.e.$ the density dependence of the symmetry energy (SE). Its precise knowledge is mandatory
for a proper understanding of nuclear structure of rare isotopes, dynamics and products of heavy-ion collisions, and
most importantly for astrophysical processes such as neutron star cooling and supernovae explosions~\cite{bar05,li08}.
Intermediate energy nuclear reactions involving radioactive beams have allowed, by studying the thickness of neutron skins,
deformation, binding energies and isospin diffusion to constrain the density dependence of SE
at densities below saturation~\cite{li98,che05}. Existing theoretical models describing its density dependence
 generally agree with each other in this density regime, but their predictions start to 
diverge well before regions with densities $\rho\geq2\rho_0$ are reached~\cite{Li:2008gp}. 

Heavy-ion collisions are the only available opportunity to recreate in laboratory fractions of nuclear 
matter with densities significantly above saturation density, presenting the opportunity for a quantitative study of the 
high-density behavior of the symmetry energy together with a few other closely related topics: in-medium 
nucleon-nucleon (NN) cross-sections, in-medium nucleon masses and the EoS of symmetric nuclear matter. These topics are 
aspects of the same, more fundamental, problem: the in-medium NN interaction. 

A quantitative study of the density dependence of SE demands the identification of observables which exhibit 
a large sensitivity to changes in its magnitude, but show small or no change with respect to variations of other 
variables or model parameters (compressibility modulus, elastic nucleon-nucleon cross-sections, etc.). 
Such observables have been a topic of intensive search during the last decade and a couple of promising
candidates have been identified: the ratio of neutron/proton ratio of squeezed out nucleons~\cite{yon07}, 
light cluster emission~\cite{che03}, $\pi^-/\pi^{+}$ multiplicity ratio in central collisions~\cite{xia09,Feng:2010zzh},
 double neutron to proton ratios of nucleon emission from isospin-asymmetric but
mass-symmetric reactions~\cite{li06b}, $N/Z$ dependence of balance energy~\cite{Sood:2010ea}
and others. The comparison of theoretical predictions for the $\pi^-/\pi^{+}$ 
multiplicity ratio with the experimental values of the FOPI collaboration~\cite{Reisdorf:2006ie} has
yielded contradictory results: the study in Ref.~\cite{xia09} points towards a soft asy-EoS at supranormal
densities, while the results of Ref.~\cite{Feng:2010zzh} favor a stiff scenario. Identifying observables that
can be compared to currently existing or in the near future available experimental data is therefore of
great importance.

This work is devoted to the study of other two observables that have been previously shown to present an important
sensitivity to changes in the isospin dependent part of EoS: elliptic flow (EF)~\cite{li96,li00,li01} and neutron-proton elliptic flow
difference (EFD)~\cite{gre03,tor07,Russotto:2011hq,Trautmann:2009kq}. Their dependence on the EoS of both symmetric and asymmetric nuclear
matter, microscopic nucleon-nucleon cross-sections and model parameters is studied in detail and exemplified
for the case of Au+Au collisions at incident energy of 400 AMeV. A comparison of our proton EF
estimates with the high precision data of the FOPI experiment is performed~\cite{and01}. Additionally, a
comparison of the theoretical values of neutron-proton EFD with the FOPI-LAND experimental 
neutron-hydrogen and neutron-proton EFD data~\cite{Leifels:1993ir,Lambrecht:1994} is presented.

The Letter is structured as follows: in Section II we review the basic ingredients of the QMD transport
model we have used and then present the pertinent details regarding the in-medium NN microscopical cross-section
and EoS of nuclear matter we have employed. We continue with Section III in which results
on proton and neutron EF and neutron-proton EFD of simulated Au+Au collisions are presented 
with an emphasis on their sensitivity to various ingredients of the transport model together with a
comparison of available FOPI and FOPI-LAND experimental data. We end with a Section
dedicated to conclusions.

\section{The transport model}

\subsection{The QMD transport model}

The heavy-ion collision dynamics is described within the framework of quantum molecular
dynamics (QMD)~\cite{aic91,kho92}, a semiclassical transport model which accounts for relevant quantum
aspects like stochastic scattering and Pauli blocking of nucleons. The T\"ubingen transport model code~\cite{uma98},
which constitutes the backbone of the code employed to produce the results reported in this work,
has been expanded by adding explicit density dependence of the microscopical nucleon-nucleon cross-sections
and building in a isospin dependent EoS. Two model parameters, the spread of the single nucleon wave function $L$
and the compressibility modulus $K$, are of relevance for the present study. There values have been set
to $2L^2$=8 fm$^2$ and $K$=210 MeV if not otherwise stated. Further details of the transport 
model can be found in Refs.~\cite{uma98,cozma2006}.

\subsection{The in-medium nucleon-nucleon interaction}

%============FIGURE 1===============================
%\begin{figure}[t]
%\begin{center}
%\begin{minipage}{0.49\textwidth}
%\epsfig{file=cs_ppnp.eps,scale=0.48} 
%\end{minipage}
%\begin{minipage}{0.49\textwidth}
%\epsfig{file=effectivemass_densdep_p025.eps,scale=0.48}
%\end{minipage}
%\caption{{\it Left Panel:} Comparison between Cugnon~\cite{cug81} and Li-Machleidt~\cite{mac93,mac94}
%parametrizations of the vacuum $NN$ cross-sections together with the in-medium parametrization at 
%density $\rho=\rho_0$ of the latter authors. {\it Right Panel:}  Effective proton and neutron masses dependence on the reduced density $\rho/\rho_0$ for 
%a few values of the asymmetry parameter $\beta$ and fixed value of the momentum $p=0.25$ GeV relative to the
%vacuum value of the nucleon mass $m_N$=0.938 GeV. The reduction factor enters the density and isospin asymmetry
%dependence of $NN$ cross-section in Eq.~(\ref{scalingcs}) (see text). 
%\label{effectivemassdensdep}}
%\end{center}
%\end{figure}

One of the important ingredients of the transport model utilized for the simulation of heavy-ion collisions are the microscopic
elastic and inelastic cross-sections. For the case of elastic collisions there are both theoretically~\cite{mac93,mac94}
and experimentally driven~\cite{li06a,gai05} studies hint for important in-medium modifications. 
For the case of total cross-sections the following density dependent parametrizations are being used
\footnote{The parametrization of $\sigma_{pp}$ we have used differs slightly from the one in~\cite{mac94}
 in order to reproduce accurately the theoretical estimates for in-medium $\sigma_{pp}$ listed
in Table I of that reference; the parametrization proposed in Eq. (1) of Ref.~\cite{mac94} is
rather inaccurate in that respect.}~\cite{mac93,mac94}
\begin{eqnarray}
\label{nncspar}
\sigma_{np}(E_{lab},\rho)&=&[ 31.5 + 0.092\times(20.2-E_{lab}^{0.53})^{2.9}] 
\times\frac{1.0+0.0034E_{lab}^{1.51}\,\rho^2}{1.0+21.55\,\rho^{1.34}} \\
\sigma_{pp}(E_{lab},\rho)&=&[23.5+0.256\times(18.501-E_{lab}^{0.52})^{3.1}] \times\frac{1.0+0.1667E_{lab}^{1.05}\,\rho^3}{1.0+9.704\,\rho^{1.2}} \nonumber
\end{eqnarray}
for energies below the pion production threshold; here $E_{lab}$ is the incident kinetic energy in MeV, while
the density $\rho$ is expressed in fm$^{-3}$. In-medium modifications of the angular dependence of the elementary
$np$ and $pp$ differential cross-sections~\cite{mac93,mac94} are neglected in the present study as they
were shown to have small impact on observables~\cite{Cozma:2009zz}. Values of the 
elementary cross-sections of reactions involving at least one excited baryon (either $\Delta$ or $N^*$) are
supposed to remain unmodified in a dense nuclear medium. Alternatively the Cugnon parametrization of 
vacuum $NN$  cross-sections~\cite{cug81} can be used. The only sizable difference between the two mentioned
parametrizations occurs in the neutron-proton channel at incident energies below 100 MeV.

The in-medium NN cross-sections should also exhibit a dependence on the isospin asymmetry
factor $\beta$. Most of existing works on the in-medium NN interactions have been focused on isospin
symmetric nuclear matter. The authors of Ref.~\cite{li05} have therefore extended the effective mass
scaling model for the in-medium NN cross-sections to isospin asymmetric nuclear matter. The central
assumption of this model is that the NN matrix elements retain their vacuum analytical expressions
in the medium, leading to a isospin asymmetry dependence indirectly through the expressions of the
effective nucleon masses as functions of $\beta$
\begin{eqnarray}
\label{scalingcs}
 \sigma_{N_1N_2}(\rho,\beta)&=&\sigma_{N_1N_2}(\rho,\beta=0)\,\frac{m_{N_1}(\rho,\beta)\,m_{N_2}(\rho,\beta)}
{m_{N_1}(\rho,\beta=0)\,m_{N_2}(\rho,\beta=0)}\,.
\end{eqnarray}

The above expression is used with minimal changes above the pion production threshold to describe both the density
and isospin asymmetry dependence of NN cross-sections by considering a dependence of nucleon mass on both
parameters.  The in-medium nucleon mass is derived from the momentum dependent part of the mean field potential 
presented in the next section by making use of the expression:
 $m_N(\rho,\beta)/m_N(\rho=0,\beta=0)=1/(1+m_N(\rho=0,\beta=0)/p\times dU/dp)$.
%In the left panel of Figure~\ref{effectivemassdensdep} the energy dependence of NN vacuum and density
%dependent cross-sections below pion production threshold is displayed for both Cugnon and Li-Machleidt
%parametrizations. In the right panel the isospin asymmetry and density dependence of in-medium nucleon masses
%is presented. Accounting for either $\rho$ or $\beta$ dependence is seen to reduce the value of NN cross-section
%at any value of the incident energy. Additionally, some difference between the vacuum Cugnon
%and Li-Machleidt parametrization is observed, especially for neutron-proton scattering 
%below 100 MeV incident energy.

\subsection{Equation of state of isospin asymmetric nuclear matter}

%================FIGURE 2===========================================
%\begin{figure}[t]
%\begin{center}
%\begin{minipage}{0.49\textwidth}
%\epsfig{file=symen_eosdensdep.eps,scale=0.45}
%\end{minipage}
%\caption{Density dependence of the symmetry energy (SE) term of the EoS for different values of the $x$ parameter.
%\label{symendensdepvarx}}
%\end{center}
%\end{figure}

The central ingredient for the study presented in this Letter is the isospin dependent part of the EoS
of nuclear matter. Theoretical studies have shown that the EoS of asymmetric nuclear matter can be approximately
expressed as $ \mathcal{E}(\rho,\beta)=\mathcal{E}(\rho)+\mathcal{E}_{sym}(\rho)\,\beta^2$
where $\beta=(\rho_n-\rho_p)/(\rho_n+\rho_p)$ is the isospin asymmetry. The value of the symmetry energy (SE), 
$\mathcal{E}_{sym}$,  is approximately known only at saturation density: $\mathcal{E}_{sym}(\rho_0$=27-36 MeV). 
For its density dependence, microscopical models lead to
divergent results at supranormal densities. We will be using a parametrization that has been developed in
Ref.~\cite{das03} starting from the Gogny effective interaction in order to obtain an explicit momentum
dependence of the symmetry energy part. It reads
\begin{eqnarray}
 U(\rho,\beta,p,\tau,x)&=&A_u(x)\frac{\rho_{\tau'}}{\rho_0}+A_l(x)\frac{\rho_{\tau}}{\rho_0}
+B(\rho/\rho_0)^\sigma(1-x\beta^2)
-8\tau x\frac{B}{\sigma+1}\frac{\rho^{\sigma-1}}{\rho_0^\sigma}\beta\rho_{\tau'} \\
&&+\frac{2C_{\tau \tau}}{\rho_0}\,\int d^3 p'\, \frac{f_\tau(\vec{r},\vec{p'})}{1+(\vec{p}-\vec{p}')^2/\Lambda^2} +\frac{2C_{\tau \tau'}}{\rho_0}\,\int d^3 p'\, \frac{f_{\tau'}(\vec{r},\vec{p'})}{1+(\vec{p}-\vec{p}')^2/\Lambda^2} \nonumber
\label{eqsympot}
\end{eqnarray}
where the parameters $A_u(x), A_l(x), B, C_{\tau,\tau'}, C_{\tau,\tau}$ and $\Lambda$ have been obtained by fitting
the momentum-dependence of $U(\rho,\beta,p,\tau,x)$ to the one predicted by the Gogny potential and requiring
that the properties of nuclear matter at saturation density are reproduced. The compressibility
modulus $K$ is set to be a soft one (210 MeV) and the value of energy symmetry at $\rho_0$ is set to 30 MeV. The parameter
$x$ can be adjusted to mimic the density dependence of the symmetry energy of various microscopical theoretical models
and has been varied in the current study between -2 and 2. The value $x=-2$ corresponds to a super-stiff asy-EoS, $x=2$ to a
super-soft asy-EoS, while setting $x=0$ the momentum independent part of SE is omitted. Most theoretical predictions
for the density dependence of SE lie in the region delimited by the $x$=-2 and $x$=1 curves~\cite{Li:2008gp}.
 Further details, including the values of  the parameters used in the calculations can be found in~\cite{li05,das03}.

%===========================FIGURE 3========================================
\begin{figure*}[tb]
\begin{center}
\begin{minipage}{0.49\textwidth}
\epsfig{file=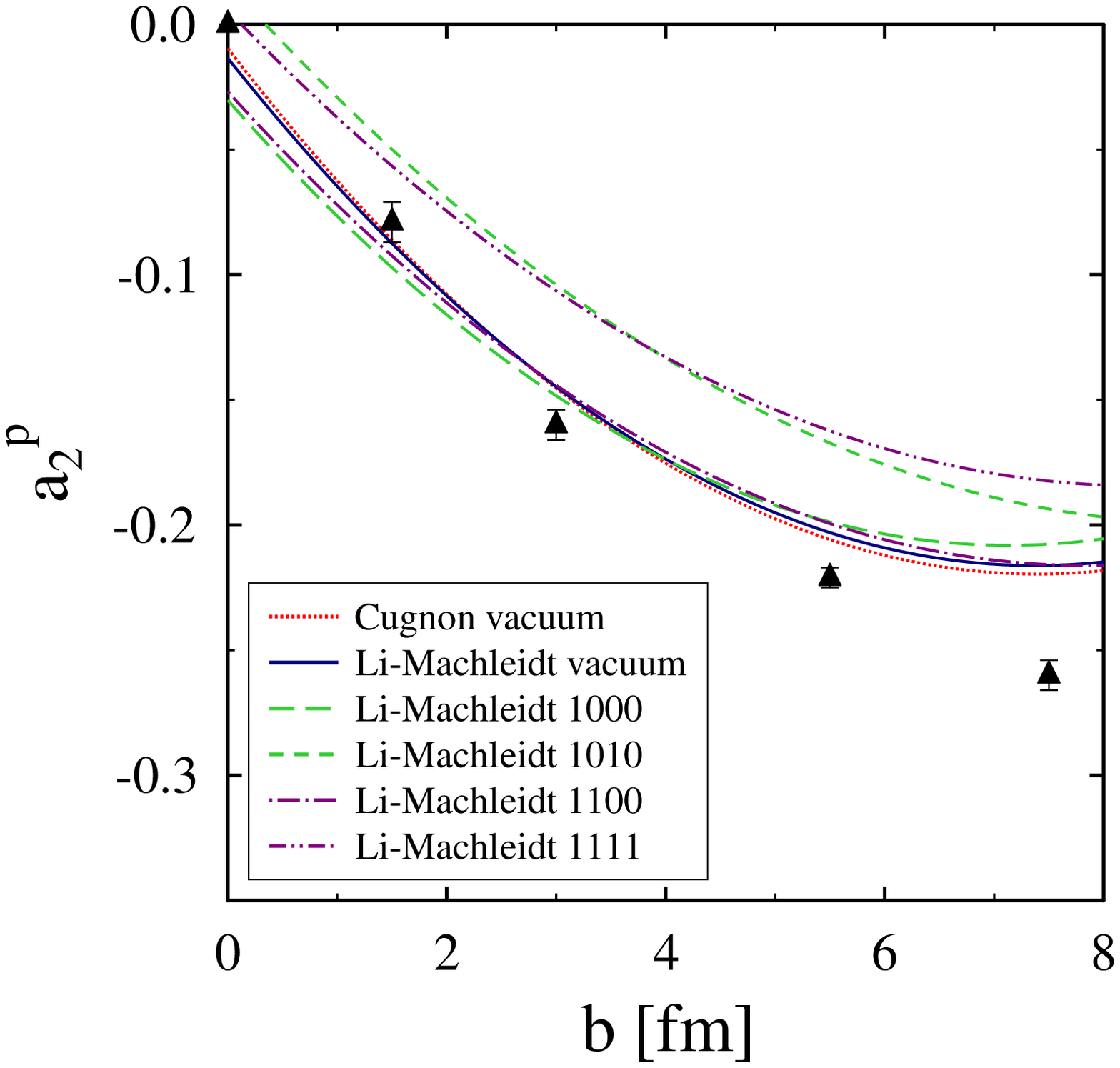,scale=0.425}
\end{minipage}
\begin{minipage}{0.49\textwidth}
\epsfig{file=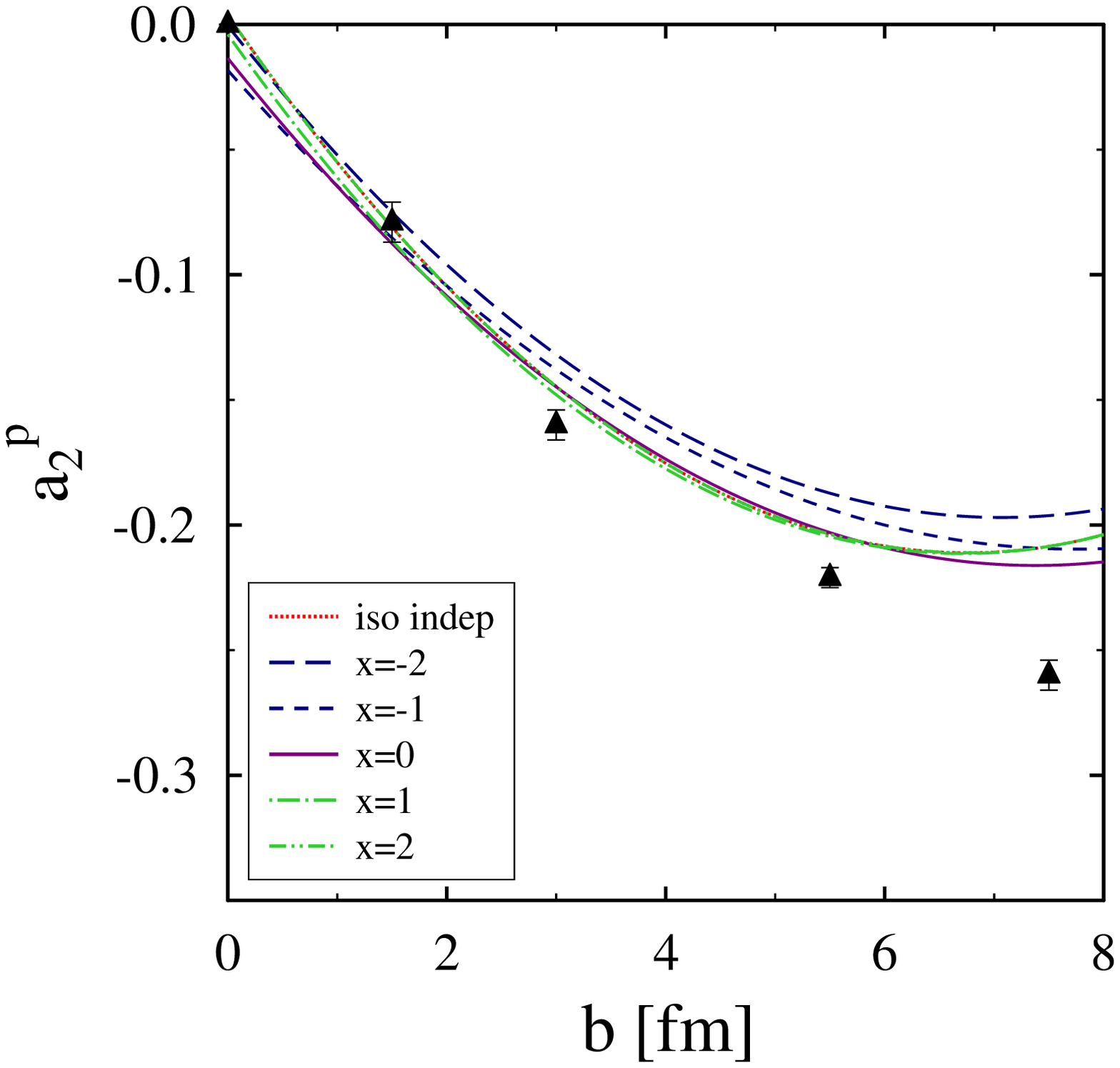,scale=0.425}
\end{minipage}
\caption{{\it Left Panel:} Dependence of the proton elliptic flow parameter $v_2^p$ in AuAu collisions at 400 AMeV
on in-medium $NN$ cross-sections. 
The isospin independent EoS is used. {\it Right Panel:} Dependence of $v_2^p$ on the asy-EoS; the vacuum Li-Machleidt $NN$
cross-sections have been used. Experimental data points are due to the FOPI collaboration~\cite{and01}.
The notation ``Li-Machleidt abcd'' stands for inclusion (1) or omission(0) 
of nucleon-nucleon cross-sections dependence on the following parameters:
$a$ - density below pion production threshold ($\pi$PT)
(parametrization in Equation~(\ref{nncspar})), $b$ - isospin asymmetry below $\pi$PT, $c$ - density above $\pi$PT and
$d$- isospin asymmetry above $\pi$PT. In the scenarios $b$, $c$ and $d$ the scaling law of Equation~(\ref{scalingcs})
has been applied. 
\label{protonflowauau400}}
\end{center}
\end{figure*}
%===========================================

%===================================FIGURE 4====================================
\begin{figure*}[tb]
\begin{center}
\begin{minipage}{0.49\textwidth}
\epsfig{file=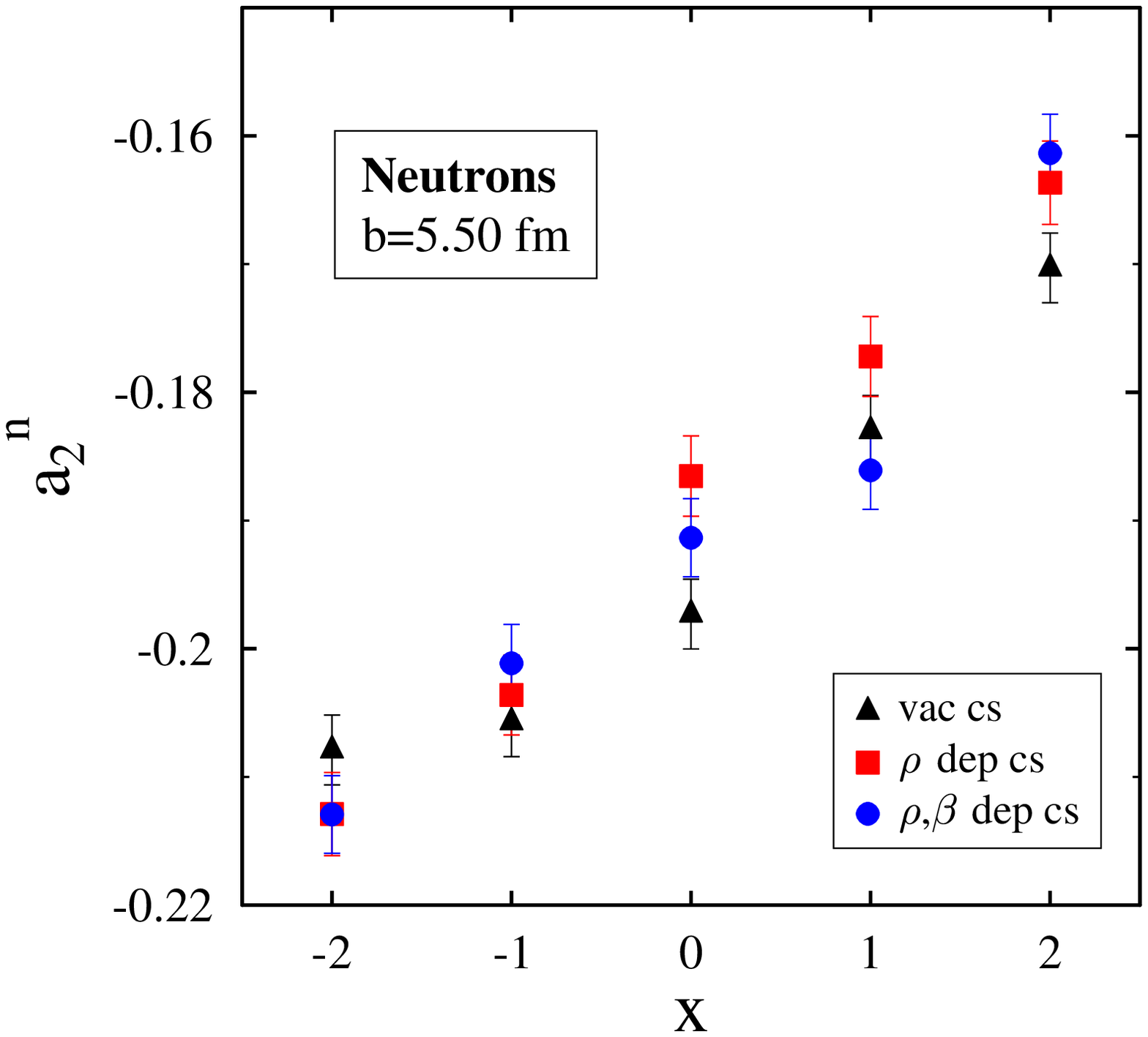,scale=0.425}
\end{minipage}
\begin{minipage}{0.49\textwidth}
\epsfig{file=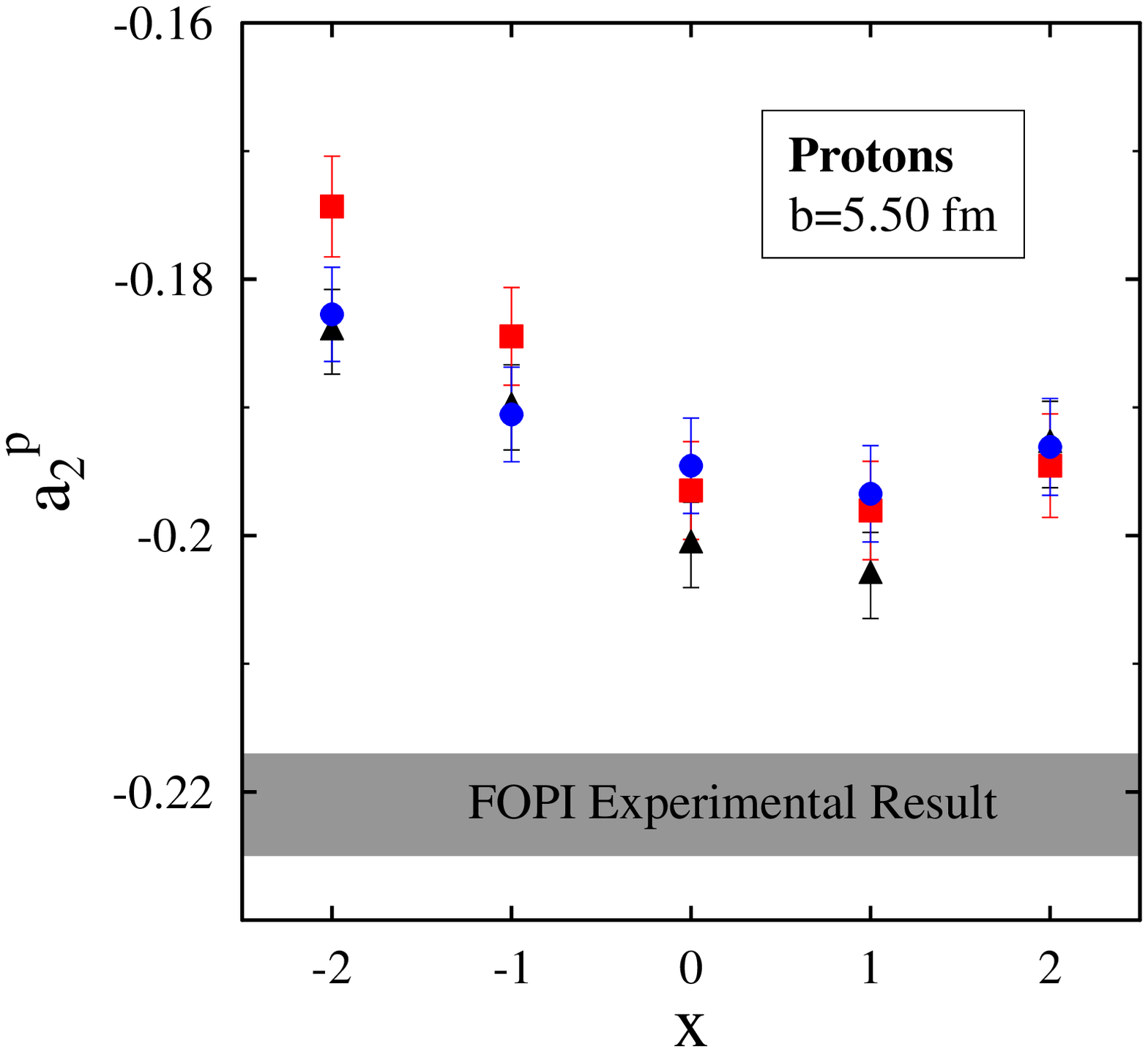,scale=0.425}
\end{minipage}
\caption{Elliptic flow of neutrons (left panel) and protons (left panel) in Au+Au collisions at
400 AMeV incident energy and impact parameter $b$=5.5 fm for given choices of the NN
microscopical cross-sections as a function of the variable $x$ that parametrizes the isospin
dependence of the asy-EoS.
\label{neutronprotonflowauau400}}
\end{center}
\end{figure*}
%===========================================

\section{Experimental observables}
\subsection{Elliptic Flow}

The azimuthal distribution of protons (or neutrons) resulted in heavy-ion collisions can be approximately described by
$ dN/d\phi=(N/(2\pi)\,\big[ \,1+v_1\,cos \phi+v_2\,cos\,2\phi\,\big]$, $v_1$ and $v_2$ being called the sidewards and
elliptic flow (EF) parameters respectively. The elliptic flow can by extracted from simulated or experimental data by
computing the following average over the respective particle specie
in the final state $v_2=(2/N)\sum_{i=1,N}\, ({p_x^i}^2-{p_y^i}^2)/{p_T^i}^2$.

We start by comparing, in Figure~\ref{protonflowauau400}, the simulated values for 
the proton EF parameter $v_2^p$ to the experimental values obtained by the FOPI collaboration
for AuAu collision at 400 AMeV incident energy. The results where obtained
by imposing the following cuts on the output particle spectra: $80^\circ \leq\phi \leq100^\circ$ and
$ 0.8 \leq p_T/p_{CM} \leq 1.8 $; additionally, the system of reference is rotated around the $y$-axis with an
angle $\theta_{flow}$~\cite{Gutbrod:1989gh} along the direction of flow.
 In the left panel the dependence of the simulated values on the NN cross-section is displayed, 
with the isospin dependence omitted. There is little difference between the two parametrizations of the vacuum
NN cross-sections, namely the ones due to Cugnon~\cite{cug81} and
Li-Machleidt~\cite{mac93,mac94} respectively. The sensitivity of $v_2^p$ to the in-medium 
NN cross-sections is moderate in size with the exception
of the case when a density dependence above $\pi$PT is considered. This is partially due to the fact that the 
flow angle $\theta_{flow}$ depends strongly on the scenario used for the in-medium NN cross-section and as a
consequence much of the EF dependence on in-medium$NN$ cross-sections is rotated away.

In the right panel of Figure~\ref{protonflowauau400} the sensitivity of $v_2^p$ to asy-EoS is studied. A monotonous
dependence on the $x$ parameters is observed with the exception of peripheral collisions in conjunction with a 
super-soft asy-EoS. A similar conclusion holds for neutrons with the exception of an opposite direction of the
monotonous variation of the EF parameter with increasing $x$. This behavior is more clearly visible in 
Fig.~\ref{neutronprotonflowauau400}where the values of neutron (left panel) and proton (right panel) EF
as a function of the parameter $x$ are plotted for mid-central collisions. We notice, additionally, 
that the difference $v_2^p(x=2)-v_2^p(x=-2)$ is in magnitude larger than the observed sensitivity on the NN
cross-sections in the left panel of Figure~\ref{protonflowauau400}.

\begin{figure*}[tb]
\begin{center}
\begin{minipage}{0.49\textwidth}
\epsfig{file=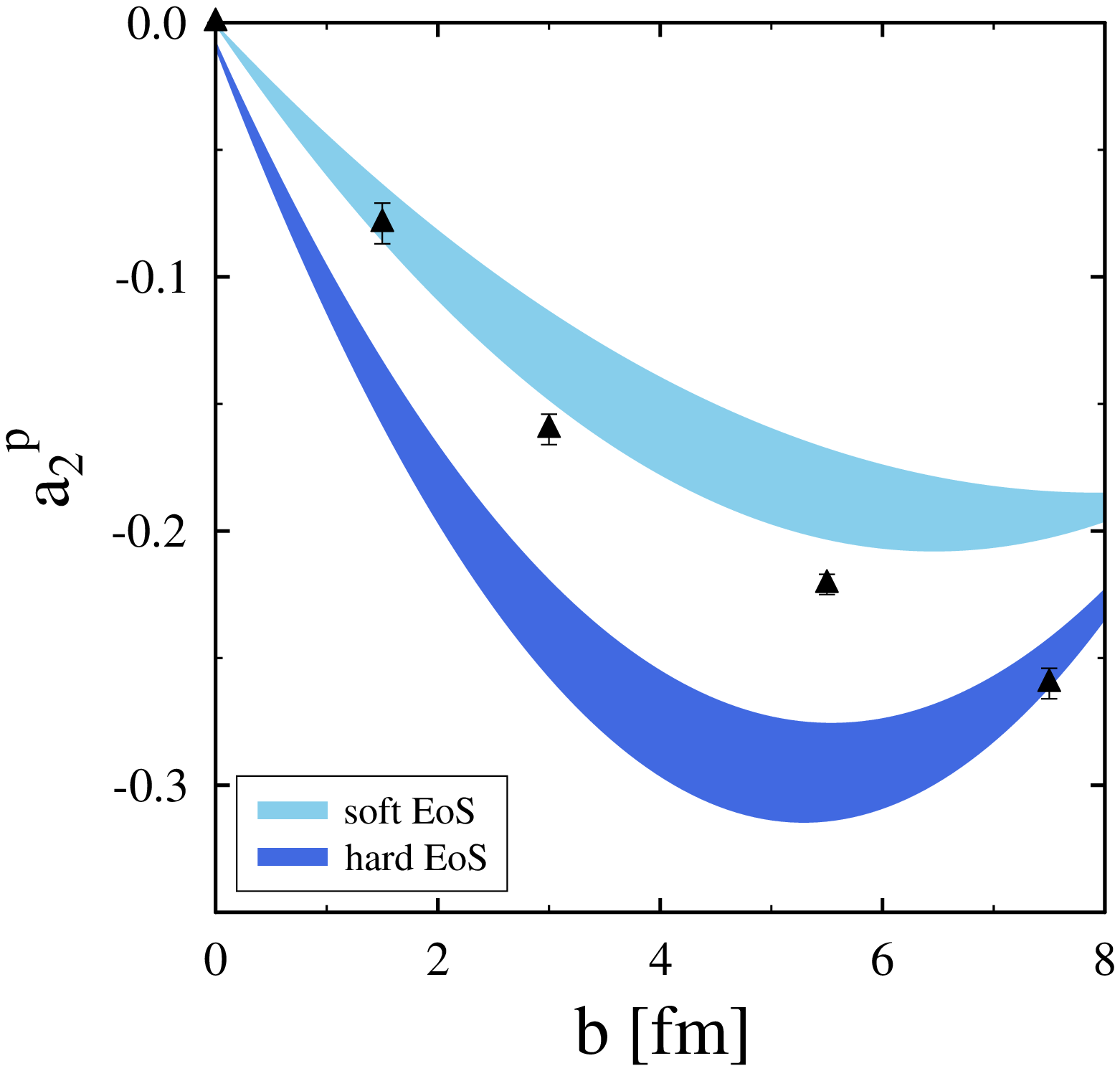,scale=0.425}
\end{minipage}
\begin{minipage}{0.49\textwidth}
\epsfig{file=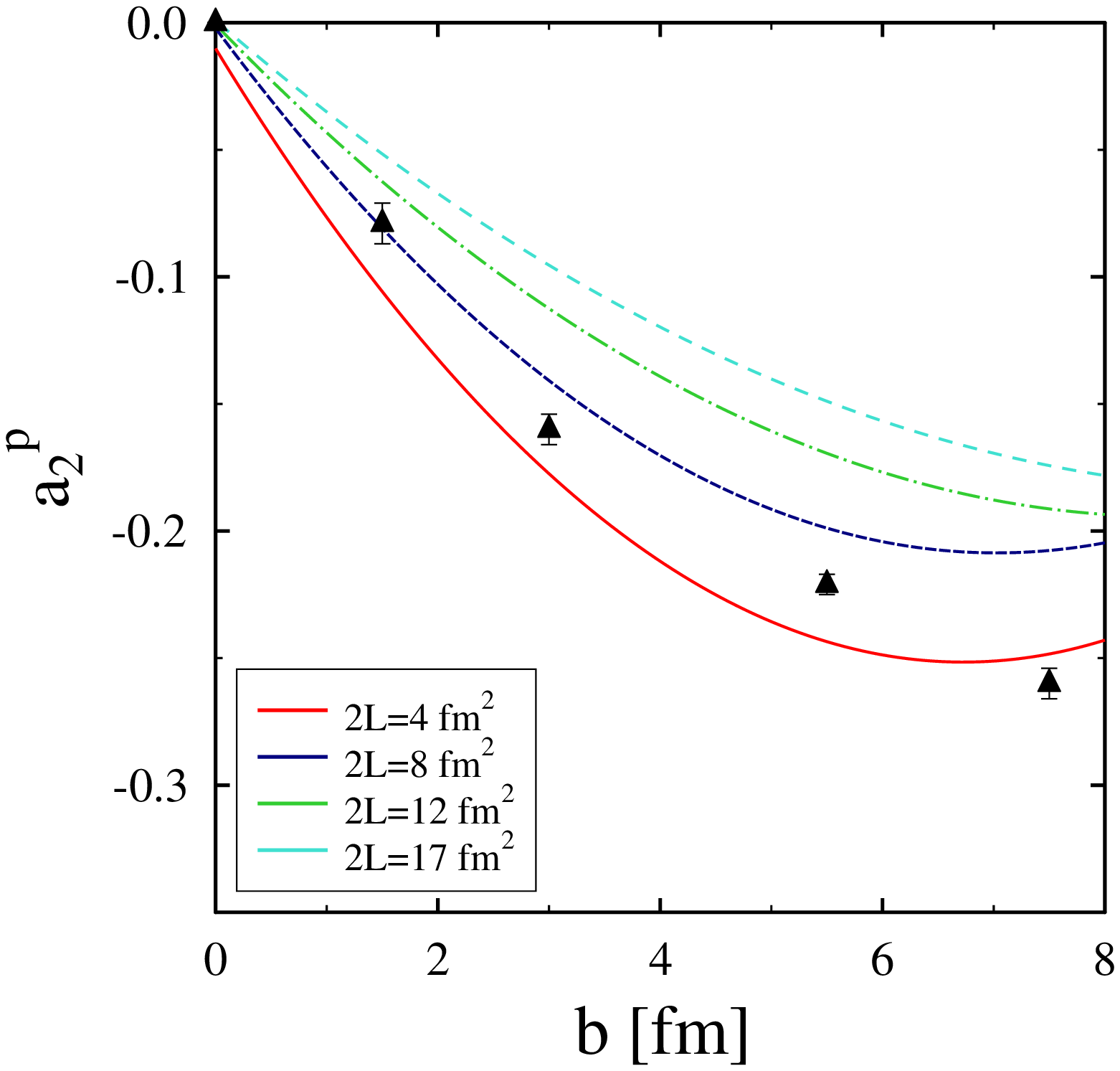,scale=0.425}
\end{minipage}
\caption{{\it Left Panel:} Sensitivity of the $a_2^p$ observable to the isospin independent EoS and various
choices of asy-EoS. Results obtained using a soft (K=210 MeV) or a hard (K=380 MeV)
are depicted by light and dark bands respectively. {\it Right Panel:} Variation of $a_2^p$ to changes
in the width of the wave packet are shown for various values of $L$ (see text).
\label{a2senseosal}}
\end{center}
\end{figure*}

The most reliable extraction of the compressibility modulus $K$ of nuclear matter has been made possible by studying
the multiplicity ratio of $K^+$ production in heavy (Au+Au) over light (C+C) nuclei at incident energies
close to 1 AGeV by the KaoS Collaboration~\cite{Sturm:2000dm,fuc01b,har06} pointing towards a soft
EoS. At lower incident energies the situation is not as clear: KaoS result points to an even softer EoS
while the study of sidewards flow by the FOPI collaboration~\cite{Andronic:2003dc} favors a soft or
hard EoS of state depending on system size. Other studies related to either sidewards flow or
elliptic flow produce similar indecisive results for collision energies of the order of a few hundreds MeV~\cite{dan02}.
In view of these results we have investigated the sensitivity of EF to changes from a soft ($K$=210 MeV)
to a hard ($K$=380 MeV) EoS. Results are shown in left panel of Fig.~\ref{a2senseosal}, the light and dark bands 
corresponding to soft and hard EoS respectively; the width of the bands were obtained by changing the asy-EoS
from a super-stiff ($x$=-2) to a super-soft ($x$=2) one. Variations in the values of EF due to changes of
the compressibility modulus $K$ are seen to exceed those due to changes in the asy-EoS scenario.

An additional indetermination of theoretical estimates is brought in by the value of the nucleon wave-packet width $L$.
Its value is usually set in literature~\cite{har98} to $2L^2$=4 fm$^2$ for light systems. For heavy systems
it is found necessary to increase the value to $2L^2$=8 fm$^2$ in order to generate nuclei with
stable static properties ($e.g.$ rms). In the right panel of Figure~\ref{a2senseosal} we display results
for proton EF for a few values of nucleon wave packet width ranging from $2L^2$=4 fm$^2$ to $2L^2$=17 fm$^2$.
Variations of the EF magnitude when changing the value of $L$ to the one customarily employed for light nuclei
or alternatively to the next larger value are sizable, of the same order of magnitude as for changes in $K$ from a soft to a hard EoS.
 
To summarize, in this section we have shown that while the proton and neutron EFs show important
sensitivity to changes in the asy-EoS, their equally large of even larger sensitivity to model
parameters like compressibility modulus $K$ or nucleon wave packet width $L$ limits their usefulness
in an attempt to constrain the supra-normal density dependence of SE in a model independent fashion.
%===================FIGURE 5=====================================
%===========================================

\subsection{Neutron-Proton Elliptic Flow Difference}

This section is dedicated to an analysis, similar to the one performed for the case of proton EF
in the previous section, of the neutron-proton elliptic flow difference (npEFD): $v_2^{n-p}=v_2^n-v_2^p$.
Due to these similarities many of the details of the last Section are identical and therefore
have been omitted where possible. The interest in this observable
stems from the following reasons. The net isospin effect to the total EoS is rather small and additionally
the pp and nn elementary cross-sections are taken to be the same. One would therefore
expect that the sensitivity to model parameters of the nucleon elliptic flow $v_2^n$ 
is similar to the one of protons. An analysis similar to the one performed for protons confirms that to be true.
In addition, the dependence of the $v_2^p$ and $v_2^n$ elliptic flows on the variable $x$ parametrizing
the asy-EoS is opposite, in line with expectations, due the fact that the neutron asy-potential is
repulsive while the proton asy-potential is attractive.
%In view of this npEFD might prove a more suitable observable for constraining
%the high density dependence of the asy-EoS.
 On the experimental side, an accurate measurement
of this observable would prove challenging as neutron distributions are rather difficult to measure accurately.

%============================FIGURE 6======================================
\begin{figure*}[tb]
\begin{center}
\begin{minipage}{0.49\textwidth}
\epsfig{file=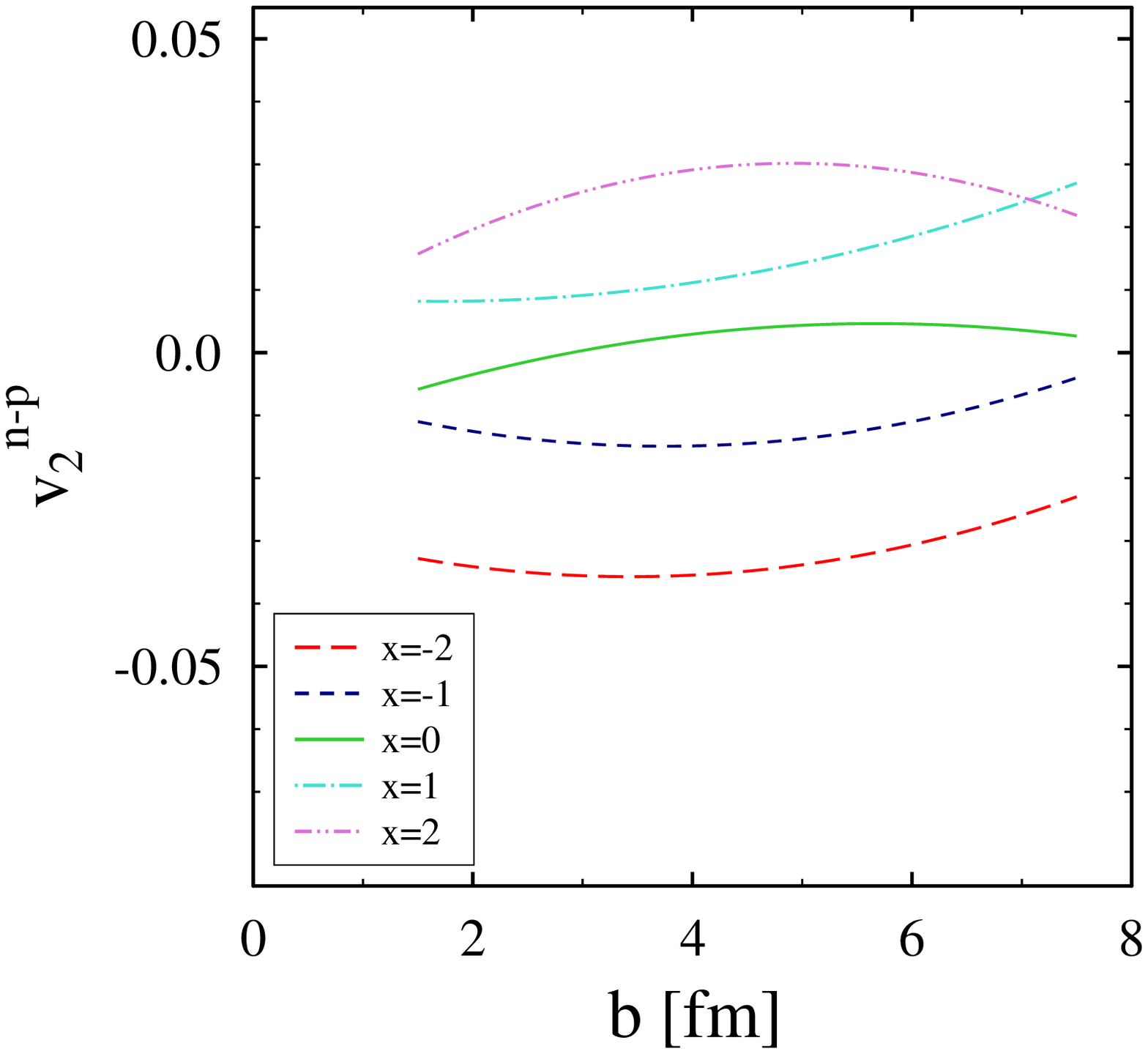,scale=0.425}
\end{minipage}
\begin{minipage}{0.49\textwidth}
\epsfig{file=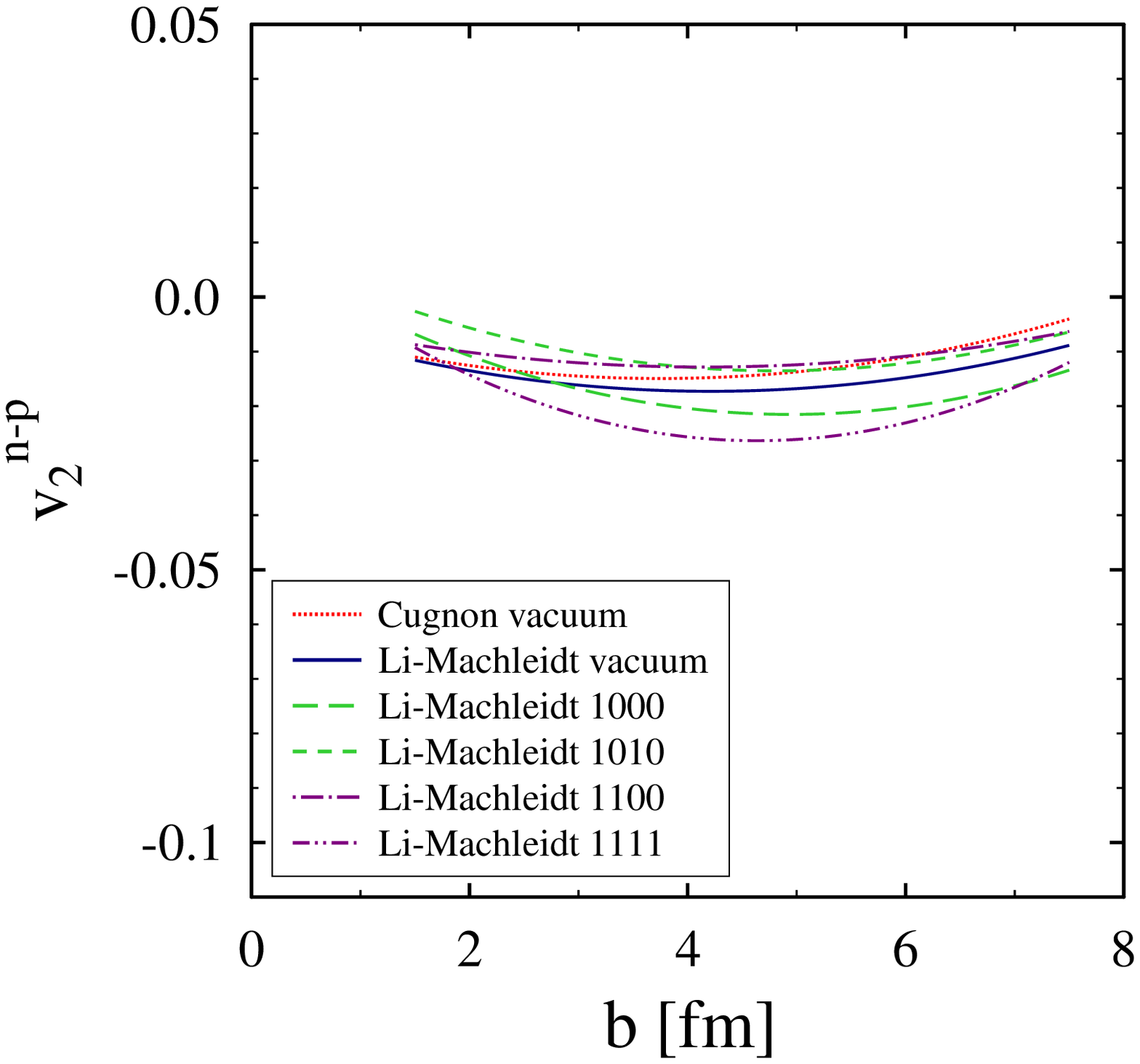,scale=0.425}
\end{minipage}
\caption{Sensitivity of npEFD to the asy-EoS ({\it left panel}) and NN in-medium cross-sections ({\it right panel}).
Statistical errors of theoretical estimates amount to about 0.005 in absolute value and have been omitted for clarity.
Labeling of curves is the same as in Fig.~\ref{protonflowauau400}.
\label{npdiffsqueezeout}}
\end{center}
\end{figure*}
%===========================================

%=============================FIGURE 7=============================================
\begin{figure*}[tb]
\begin{center}
\begin{minipage}{0.49\textwidth}
\epsfig{file=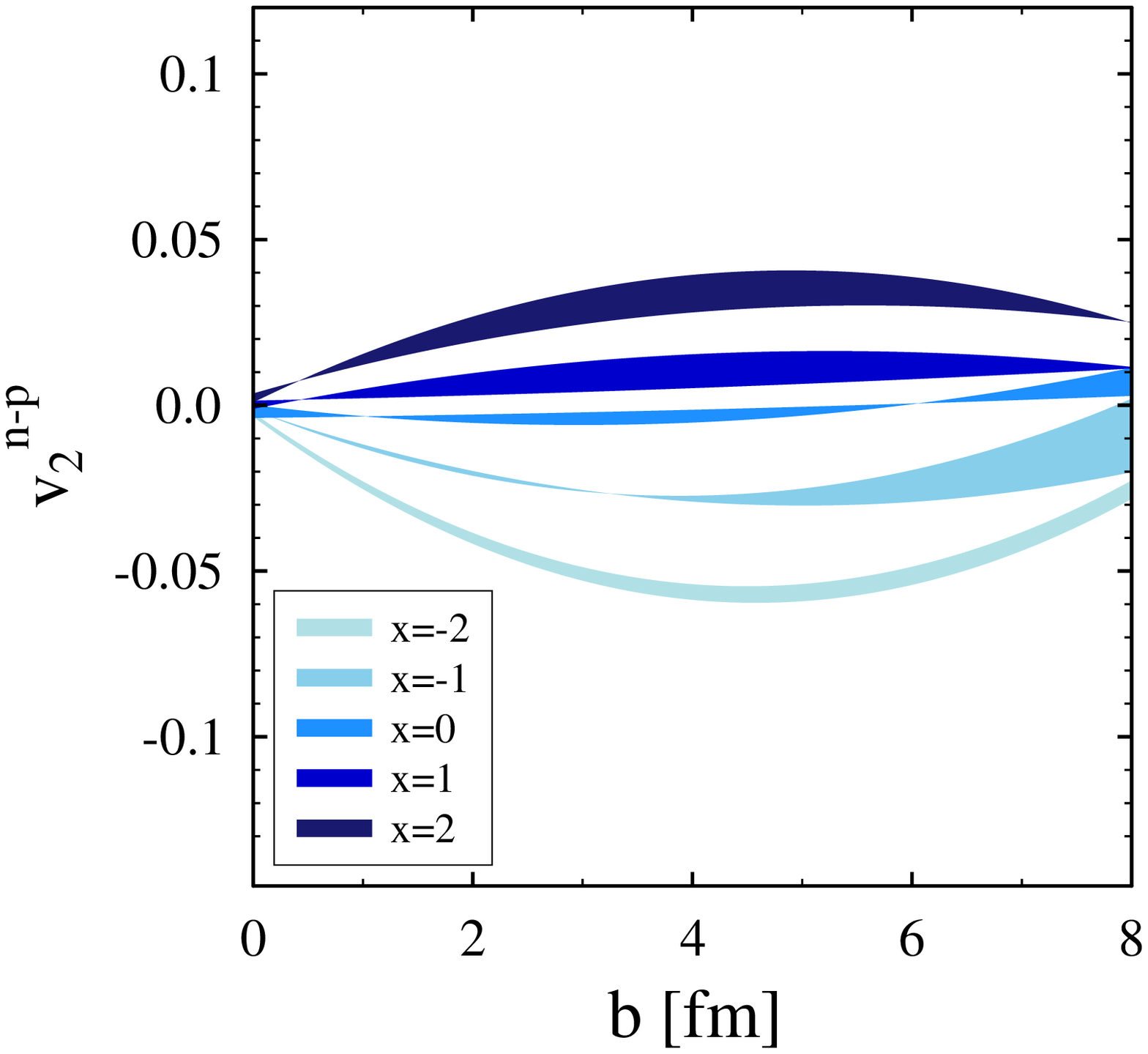,scale=0.425}
\end{minipage}
\begin{minipage}{0.49\textwidth}
\epsfig{file=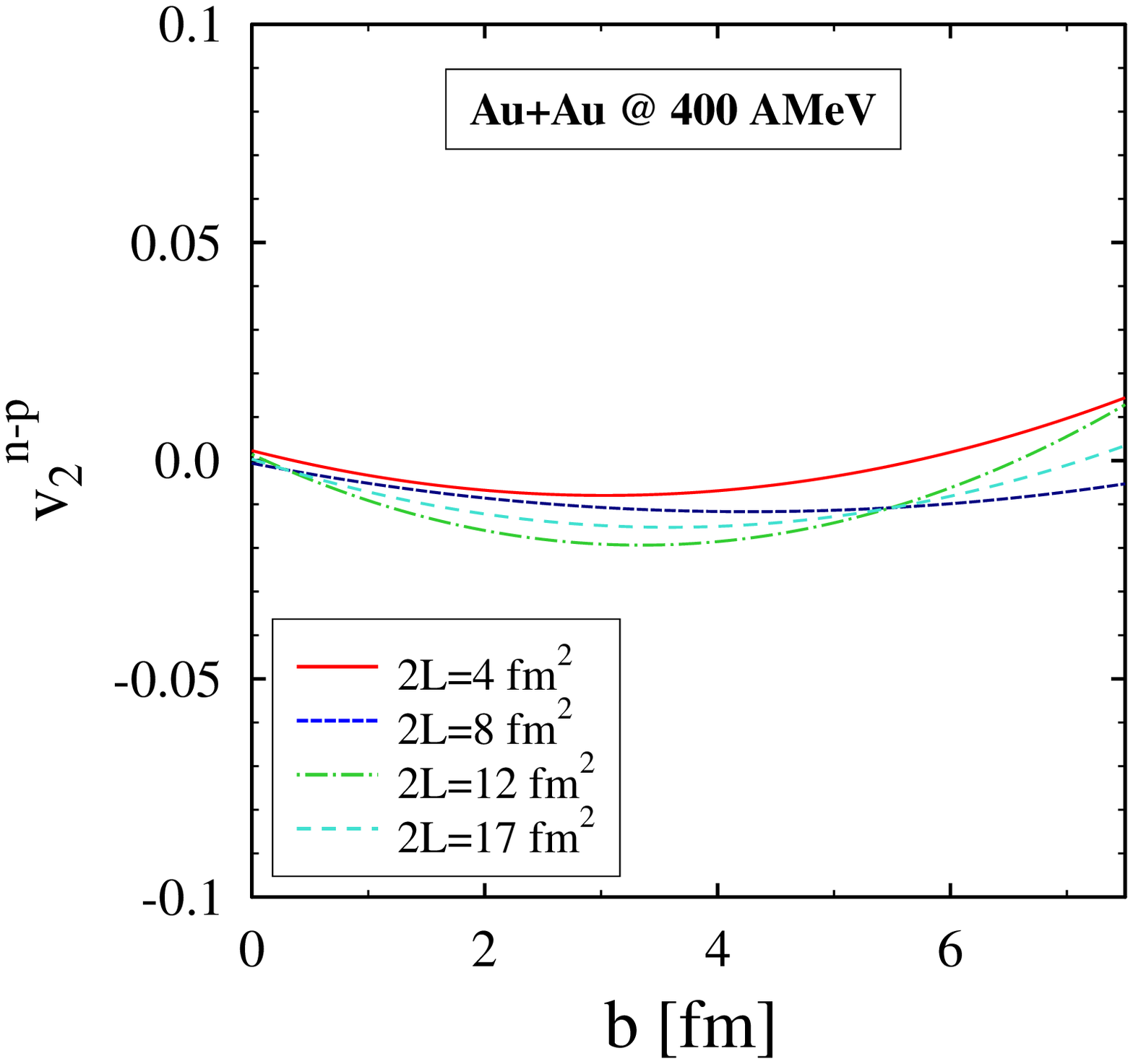,scale=0.425}
\end{minipage}
\caption{The same as Fig.~\ref{a2senseosal} but for the neutron-proton elliptic flow difference.
\label{diffa2senseosal}}
\end{center}
\end{figure*}
%===========================================

%===============================FIGURE 8==============================================
\begin{figure}[h]
\begin{center}
\epsfig{file=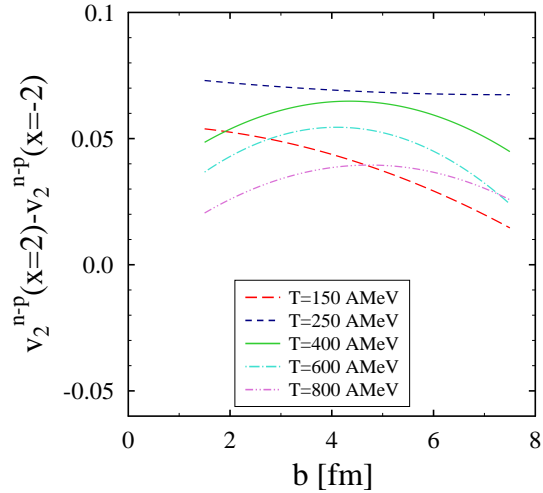, scale=0.425}
\caption{Energy and impact parameter dependence of the splitting asy supersoft-superstiff of npEFD. }
\label{npdiffsqueezendep}
\end{center}
\end{figure}
%===============================================================

%===============================FIGURE 9=============================================
\begin{figure}[h]
\begin{center}
\epsfig{file=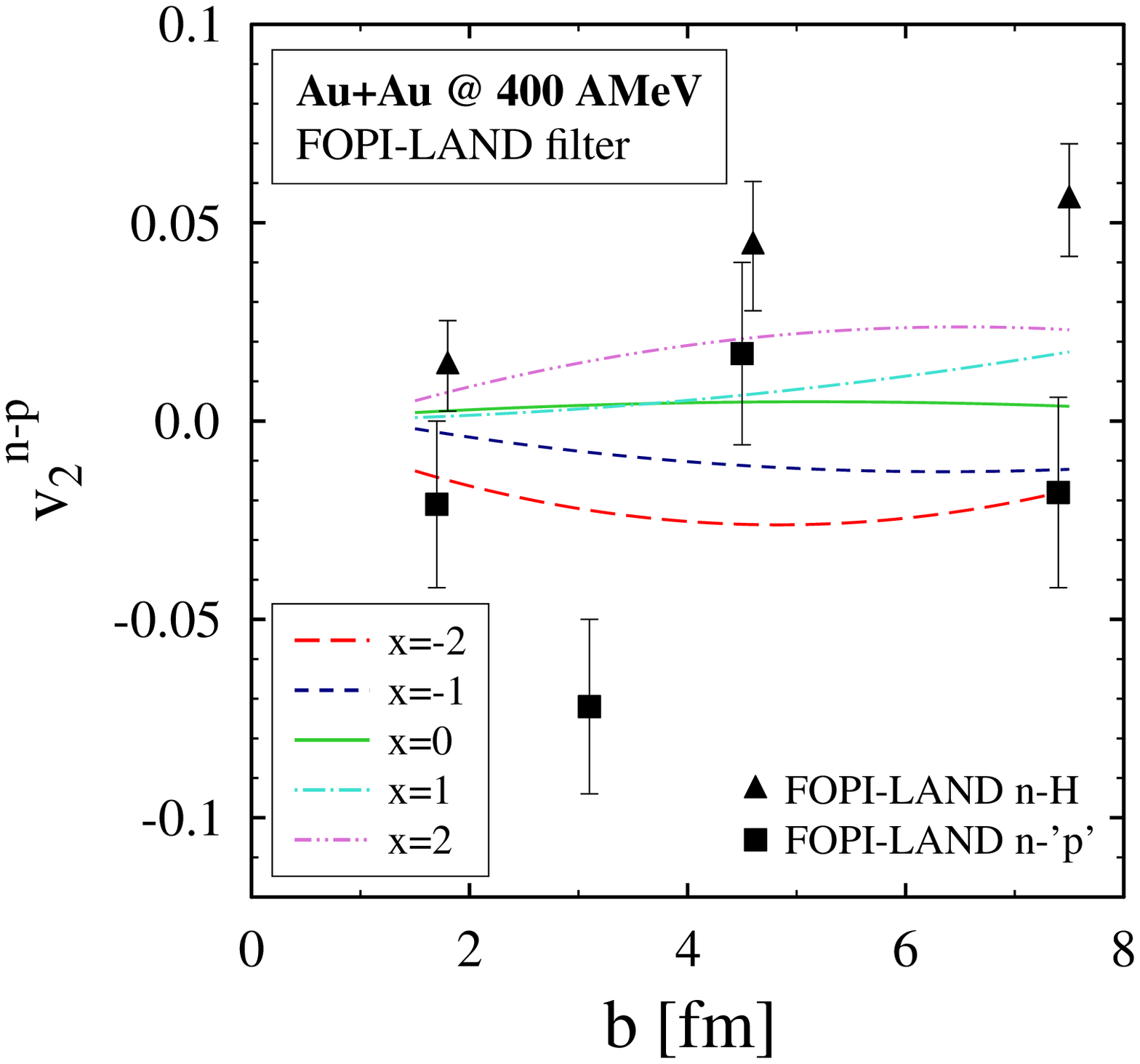, scale=0.425} 
\caption{Comparison between theoretical estimates of npEFD and the FOPI-LAND experimental
data for nHEFD~\cite{Leifels:1993ir} and n'p'EFD~\cite{Lambrecht:1994}.}
\label{npefdfopiland}
\end{center}
\end{figure}
%=========================================

In Figure~(\ref{npdiffsqueezeout}) a similar analysis as the one performed for protons in 
Figure~\ref{protonflowauau400} but for npEFD 
is presented. The left side panel presents the dependence of this observable on the choice of the
asy-EoS. The splitting between the extreme cases, $x=-2$ and $x=2$, is constant as a function of the
impact parameter and its magnitude represents about 30-40$\%$ of that of either neutron or proton
EF the same energy and impact parameter. Additionally, the sensitivity to in-medium $NN$
nucleon-nucleon cross-sections is almost an order of magnitude smaller (see right panel of
Figure~(\ref{npdiffsqueezeout})).

A study of the dependence on model parameters like compressibility
modulus and width of the Gaussian wave packet shows a sizable suppression as compared to
the proton or neutron EF case, see Fig.~\ref{diffa2senseosal}. The width of the bands in the left
panel of Fig.~\ref{diffa2senseosal} represents the sensitivity of npEFD to changes of the isospin independent
EoS from a soft to hard one. The results for various choices of variable $x$ are clearly disentangled from 
each another. Complementary, variations of npEFD due to changes in $L$, the nucleon wave packet width,
are also small, of the order of the statistical error of the presented results.

The results presented in Fig.~\ref{npdiffsqueezeout}) and~\ref{diffa2senseosal} for npEFD recommend it
as a viable observable for constraining the high density behavior of the asy-EoS in a model independent way. 
The results presented in these figures were obtained by simulating Au+Au collisions at an incident energy of 400 AMeV and
applying the FOPI filter. In Figure~(\ref{npdiffsqueezendep}) we extend that study 
by varying the impact energy from 150 AMeV to 800 AMeV together with a variable impact parameter.
We plot the impact parameter dependence of $\Delta(npEFD)$=npEFD$[x=-2]$-npEFD$[x=2]$ for various colliding energies,
a close to zero value for this variable signaling no sensitivity of npEFD to asy-EoS.  The most favorable energy domain
stretches from T=250 AMeV to almost T=600 AMeV and as energy increases in this interval the impact parameter window
shrinks to mid-central collisions. For incident energies larger than 600 AMeV only mid-central collisions 
show important sensitivity of npEFD to the asy-EoS scenario but with decreasing magnitude as energy increases.
The magnitude of the of the observed effect demands, for a successful constraining of the asy-EoS, experimental data
for the elliptic flow of both charged and neutral particles of high quality, of an accuracy of 1\% or better.

We have saved for last a comparison of our estimates for npEFD with the experimental values of
neutron-hydrogen EFD (nHEFD)~\cite{Leifels:1993ir} and neutron-'proton' EFD (n'p'EFD)~\cite{Lambrecht:1994}
obtained by the FOPI-LAND collaboration. In addition
to charged particles, the FOPI-LAND experiment has succeeded to measure azimuthal distributions
of neutrons, providing experimental evidence of their elliptic flow. Results for charged particles
with $Z$=1 (hydrogen) have also been published in Ref.~\cite{Leifels:1993ir}. Extracting pure proton distributions
is not straightforward as the calorimeter resolution of the FOPI-LAND experimental setup is insufficient
to separate the H isotopes. In Ref.~\cite{Lambrecht:1994} conservative cuts, that leave
out the tail contributions of proton energy spectra while retaining some deuteron events, 
have been applied to extract proton spectra (denoted here as 'proton' or 'p' in order to stress the difference
from a pure proton spectrum). The FOPI-LAND experimental values for proton
elliptic flow should therefore be considered as rough estimates only, and due to the controversial method
employed to extract them a comparison with theoretical estimates should be viewed with care.

It is well known that flow of heavier isobars is stronger and as a result the values of $v_2^H$ represent upper values
(in absolute magnitude) for $v_2^p$, nHEFD being therefore an upper value for npEFD. 
The comparison of our estimates for npEFD with the FOPI-LAND nHEFD  and n'p'EFT experimental 
results is presented in Figure~(\ref{npefdfopiland}), providing an upper limit to the softness of  the asy-EoS.
A comparison with theoretical estimates of nHEFD should provide tighter constraints; work in that direction is
in progress. The n'p'EFD experimental data show a scattered
dependence on the impact parameter most likely due to the procedure employed to eliminate deuteron and triton events
from experimental spectra.  They are shown here in an attempt to
stress that further refinements on the experimental side are necessary for a meaningful comparison of this observable to
theoretical models. The ASY-EOS collaboration, a follow-up of the LAND, FOPI and CHIMERA
collaborations, is planning  to measure both proton and neutron azimuthal distributions
with an increased accuracy which will have the potential to constrain of the high density dependence of
asy-EoS through observables like npEFD.

\section{Conclusions}

We have employed an isospin dependent QMD transport model to study the dependence of
elliptic flow observables on the asy-EoS. The proton and neutron elliptic flows show a monotonous
dependence on the variable $x$ that parametrizes the various asy-EoS employed, but in opposite directions
for protons ans neutrons respectively. A sizable dependence of these observable on model
parameters like the compressibility modulus $K$, the
width of the Gaussian wave packet describing particles in QMD, $L$, and to a lesser extent to in-medium effects
on the microscopic NN cross-sections renders them useless for constraining, in a model independent way,
the high density dependence of asy-EoS. 

The neutron-proton elliptic flow difference shows (npEFD), in contrast,
a sizable dependence on the asy-EoS and a almost an order of magnitude smaller dependence on model
parameters and NN in-medium cross-sections. A study of the dependence of its magnitude 
on collision energy and impact parameter leads to the conclusion that neutron-proton differential squeeze-out
can be effectively used to constrain the asy-EoS when measured in mid-central heavy-ion collisions at 
incident beam energies between 250-600 MeV. Nevertheless, an effective constraint of the asy-EoS would require
experimental values for the elliptic flow of both charged and neutral particles of an accuracy of 1\% or better.

A comparison with existing experimental data for the
neutron-hydrogen elliptic flow difference, due to the FOPI-LAND collaboration, sets and upper limit on the
softness of the asy-EoS. The experimental values for the neutron-proton elliptic flow difference show a
scattered dependence on the impact parameter, most likely due to the controversial method used in the
analysis of the experimental data, and as such a comparison with theoretical estimates is unreliable and
should be viewed with care. The ASY-EOS Collaboration promises to remedy these shortcomings and together
with a higher accuracy measurement of neutron and proton elliptic flows will potentially present the opportunity to
constrain the high dependence of asy-EoS through observables like neutron-proton elliptic flow difference. 

\section{Acknowledgments}
M.D.C. acknowledges partial financial support from the Romanian Ministry of Education and Research
through CNCSIS grant RP9. Numerical simulations have been performed on the computing cluster
of the Hadronic Physics Departement of IFIN-HH.

%Proton and deuteron rapidity distributions and nuclear stopping in 96Ru(96Zr)+96Ru(96Zr) collisions at 400AMeV
%\bibitem{hon02} B. Hong, $\it et\,al.$ (FOPI Collaboration), \Journal{\PRC}{66}{034901}{2002};

%equation of state for dense nuclear matter (Argonne v14. Urbana v14 + other models with also 3N forces)
%\bibitem{wir88} R.B. Wiringa, V. Fiks, and A. Fabrocini, \Journal{\PRC}{38}{1010}{1988}

%new parametrizations of the optical potential
%\bibitem{har94} C. Hartnack, and J. Aichelin, \Journal {\PRC}{49}{2801}{1994}

%Equation of State of Asymmetric Nuclear Matter and Collisions of Neutron-Rich Nuclei
%\bibitem{li97} B.-A. Li, C.M. Ko, and Z. Ren, \Journal{\PRL}{78}{1644}{1997}.

%isospin dependence of collective flow in heavy-ion collisions at intermediate energies
%\bibitem{li04} B.-A. Li, C.B. Das, S. Das Gupta, and C. Gale, \Journal{\PRC}{69}{011603R}{2004}

%momentum dependence of symmetry potential in asymmetric nuclear matter for transport model calculations
%\bibitem{das03} C.B. Das, S. Das Gupta, C. Gale, and B.-A. Li, \Journal{\PRC}{67}{034611}{2003}

%Evidence for collective expansion in light particle emission following Au+Au collisions at 100,150 and 250 AMeV
%\bibitem{pog95} G. Poggi {\it et al.}, \Journal{\NPA}{586}{755}{1995};

\end{document}